\documentclass[letterpaper]{article}
\usepackage[T1]{fontenc}

\usepackage{geometry}
\usepackage{setspace}

\usepackage[style = chem-acs]{biblatex}
\usepackage{graphicx}
\usepackage{float}
\newfloat{scheme}{htbp}{los}
\floatname{scheme}{Scheme}
\floatname{chart}{Chart}
\newfloat{graph}{htbp}{loh}

\usepackage{amssymb,amsmath}
\usepackage{hyperref}

\usepackage{braket}
\usepackage{array}
\usepackage{booktabs}
\usepackage{threeparttable}
\usepackage{bm}
\usepackage{epstopdf}
\usepackage[version=4]{mhchem}
\usepackage{multirow}
\usepackage{siunitx}
\newrobustcmd\B{\DeclareFontSeriesDefault[rm]{bf}{b}\bfseries}

\hypersetup{
    bookmarks=true,         
    unicode=false,          
    pdftoolbar=true,        
    pdfmenubar=true,        
    pdffitwindow=false,     
    pdfstartview={FitH},    
    colorlinks=false,       
    linkcolor=black,          
}
\graphicspath{{figures/}}

\usepackage{authblk}
\author[1]{Jiří Suchan*}
\author[2,$\dagger$]{B. Scott Fales}
\author[1,3]{Benjamin G. Levine*}
\author[4]{Eva Muchová}
\author[4]{Petr Slavíček*}
\affil[1]{Institute of Advanced Computational Science, Stony Brook University, Stony Brook, New York 11794, United States}
\affil[2]{Department of Chemistry, Michigan State University, East Lansing, Michigan 48824, United States}
\affil[$\dagger$]{Current address: NVidia, Santa Clara, California 95051, United States}
\affil[3]{Department of Chemistry, 
Stony Brook University, Stony Brook, New York 11794, United States}
\affil[4]{Department of Physical Chemistry, University of Chemistry and Technology, Prague, Technická 5, 166 28 Prague, Czech Republic}
\title{Real-Time Emergence of Charge-Transfer-to-Solvent States from Core Excitation}
\date{*Email: jiri.suchan@stonybrook.edu, ben.levine@stonybrook.edu, petr.slavicek@vscht.cz}

\begin{document}

\maketitle

\begin{abstract}
Charge-transfer-to-solvent (CTTS) excitations provide a chemically central route to generating hydrated electrons and initiating redox chemistry in solution, yet the earliest stage of CTTS---the formation of the excited state itself---is usually treated as instantaneous. Here we present a time-domain perspective of how CTTS character builds up during \emph{core-level} photoexcitation of an aqueous metal ion. Using time-dependent configuration interaction, we simulate the coherent evolution of a dense manifold of core-excited states and track the ultrafast flow of electronic charge from the initially localized site into solvent-supported final states. We find that the dynamics evolves from a few-state, oscillatory behavior to effectively irreversible delocalization, as the charge disperses among many coupled configurations, providing a microscopic mechanism for the early-time emergence of CTTS character. Our results offer a transparent real-time interpretation of what core-level spectroscopies and core-hole-clock-type measurements can probe in solutions, outlining experimental signatures for probing the build-up of CTTS states on the core-hole-lifetime timescale. Looking ahead, attosecond and sub-femtosecond X-ray pump--probe approaches at X-ray free-electron lasers provide a realistic route to directly time-resolve the core-excited CTTS wave packets in solution.

\end{abstract}

\maketitle

\section{Introduction}

Attosecond science has opened experimental access to electronic motion on its natural timescale, offering a more direct connection between real-space charge dynamics and the “electron-flow” concepts that underpin mechanistic reasoning in chemistry.\cite{scudder2023electron,nisoli2017attosecond} To date, most molecular attosecond studies have focused on charge migration initiated by sudden ionization, where coherent oscillations of charge density are observed.\cite{calegari2014ultrafast} While these remarkable high-energy nonequilibrium experiments have provided fundamental insight into ultrafast electron dynamics, their connection to the concepts of electron flow as used in chemistry is indirect. It is therefore natural to ask which attosecond electronic processes are most closely tied to the formation of excited states and charge-transfer pathways that matter in chemistry.


We focus here on electron flow during photoexcitation. In chemistry, excitation is typically treated as an instantaneous Franck--Condon (vertical) step, and the formation of the excited state is therefore rarely viewed as a dynamical process. As a result, the earliest stages of excitation often remain obscured. Using time-dependent electronic structure theory, we show how charge-transfer character can emerge from coherent superpositions, interference among many electronic configurations, and rapid environment-assisted delocalization. We further discuss how this early-time formation can be visualized within a time-resolved quantum-mechanical description and, in principle, probed experimentally in a core-excited charge-transfer-to-solvent (CTTS) process.

\begin{figure}[h!]
\centering
\includegraphics[width=.65\textwidth]{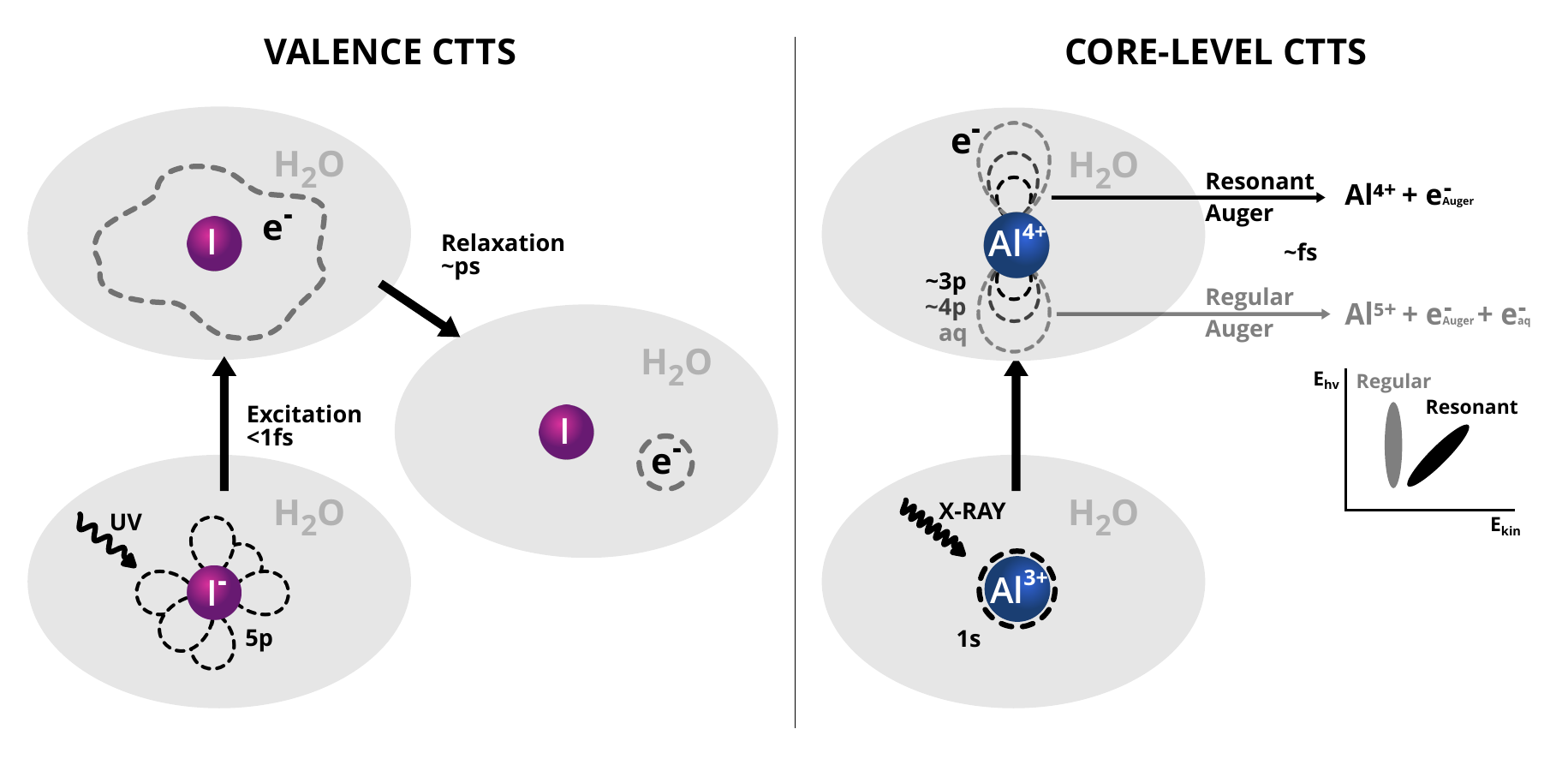}
\caption{Scheme of various CTTS processes. Valence excitation in I$^{-}$ and subsequent nuclear relaxation that leads to formation of $e_{\mathrm{aq}}^-$ on the left. Core-excitation of Al$^{3+}$ and subsequent Auger processes occurring on the femtosecond timescale are shown on the right.}%
\label{fig:scheme}
\end{figure}

The formation of excited states in condensed-phase environments is the natural entry point to photochemistry, radiation chemistry, and redox chemistry. We investigate electron transfer into the solvent leading ultimately to the formation of the hydrated electron, $e_{\mathrm{aq}}^-$.\cite{herbert_hydrated_electron_2017} As depicted on the left-hand side of Fig.~\ref{fig:scheme}, this process starts with the formation of optically bright CTTS states,\cite{rabinowitch_electron_transfer_spectra_1942, carterfenk_birth_hydrated_electron_2023} in which an electron is promoted from a solute into diffuse, solvent-supported electronic states. In this initial sub-fs stage, nuclear motion is not yet dynamically important. On longer timescales (fs to ps), however, the initially excited states relax toward the hydrated electron $e_{\mathrm{aq}}^-$ via nuclear rearrangement and non-adiabatic processes. This later, nuclear-driven relaxation from valence CTTS states of, e.g., halides and pseudohalides such as I$^-$ or SCN$^-$ has been extensively investigated using ultrafast optical spectroscopy, time-resolved photoelectron spectroscopy of liquid microjets and clusters, and mixed quantum--classical simulations.\cite{messina_real_time_ctts_2013,barthel_sodide_ctts_2003,kothe_ctts_dynamics_iodide_2015,kammrath_ctts_precursor_clusters_2005,okuyama_ctts_iodide_trpes_2016,lubcke_trpes_solvated_electron_2010} These studies have established the dynamics following the creation of locally excited "trapped" CTTS states: solvent-driven evolution first yields more delocalized pre-hydrated states, followed by the formation of the equilibrated hydrated electron on the 100--400~fs timescale. In the present work, by contrast, we focus on the earlier electronic stage---the build-up of CTTS character itself---and on its core-excited counterpart shown on the right-hand side of Fig.~\ref{fig:scheme}.

Charge transfer into the solvent can also be initiated by excitation of core electrons with soft X-rays. This strategy offers several advantages: it launches the process with element and site specificity, and the finite core-hole lifetime provides an intrinsic ultrafast timescale that can be related to excited-state formation.\cite{ottosson_core_hole_clock_review_2012} Resonant core-level excitation followed by Auger decay has therefore been used in aqueous electrolytes and at interfaces to monitor attosecond--femtosecond charge delocalization within the core-hole clock (CHC) framework.\cite{bruehwiler_core_spectroscopies_review_2002,fohlisch2005direct} In the present context, CHC serves primarily as an experimentally relevant point of contact, rather than as the central theoretical focus of this work.

Recent solution-phase studies have begun to extend core-level CTTS concepts beyond surfaces and weakly coupled interfaces, using core-level absorption and resonant Auger decay as a site-selective probe of electron delocalization in the presence of a finite core-hole lifetime. In particular, for hydrated Na$^+$, Mg$^{2+}$, and Al$^{3+}$, combined liquid-jet photoemission, $1s$-edge X-ray absorption, and Auger spectroscopy revealed dense core-excited manifolds with strongly mixed metal--water character and signatures consistent with ultrafast redistribution of the excited electron into the hydration environment.\cite{muchova_coreholeclock_ctts_2024} A photon-energy-dependent analysis of the Auger spectra yielded apparent delocalization times ranging from a few hundred attoseconds near threshold to a few tens of attoseconds at higher excitation energies, indicating that core-excited CTTS-like resonances can persist even above the nominal ionization threshold.\cite{muchova_coreholeclock_ctts_2024} These observations place the underlying electron dynamics on sub-femtosecond to femtosecond timescales comparable to the core-hole lifetime, making a real-time description of the evolving core-excited wave packet both natural and timely. They also point toward a regime that can increasingly be connected not only to CHC-type measurements, but also to emerging direct time-resolved X-ray experiments.

Here, we therefore apply a fully quantum-mechanical real-time description of charge transfer into the solvent. Real-time electronic-structure methods,\cite{goings2018real,li2020real} including real-time time-dependent density functional theory (TDDFT),\cite{herbert_zhu_xray_tddft_2023} algebraic diagrammatic construction (ADC),\cite{lunnemann2009ultrafast,ruberti2023advances} and various flavors of time-dependent configuration interaction (TDCI),\cite{greenman_tdcis_strongfield_2010,carlstrom_tdcis_atomic_2022,carlstrom_tdcis_molecular_2022,schriber_adaptive_ci_2019,pabst_tdci_2012,kochetov2021rhodyn} have already proven powerful for describing ultrafast charge migration, charge transfer, and nonlinear spectroscopy in systems with dense manifolds of excited states. Explicit treatment of deep core levels, however, remains challenging for TDDFT\cite{besley2021modeling} because of self-interaction, approximate exchange--correlation kernels, and the frequent use of pseudopotentials that replace explicit core electrons. This motivates the use of an \textit{ab initio} TDCI approach based on multireference configuration-interaction (CI) wavefunctions, which can describe core--valence correlation and strongly mixed core-excited states on equal footing.

We employ time-dependent configuration interaction built on multireference CIS (MRCIS) to study CTTS-state formation following Al$^{3+}$ $1s$ excitation in water. Our focus is the coherent electronic dynamics before Auger decay in a finite-size cluster, which we interpret as the initial build-up of CTTS character and electron delocalization. The choice of Al$^{3+}$ offers two key advantages. First, the initial $1s$ orbital is a compact and well-defined $s$ state, so within the dipole approximation the excited electron is initially promoted into a $p$-symmetry wave, which simplifies the interpretation of the emerging density patterns and their solvent-induced scattering. Second, the $1s$ core-excited states of hydrated Al$^{3+}$ are metastable, with Auger-limited lifetimes on the order of 1--2~fs, comparable to the timescale on which the core-excited wave packet reorganizes within the first and second solvation shells. This makes hydrated Al$^{3+}$ a particularly suitable system for connecting real-time electronic dynamics to core-level observables on the core-hole-lifetime timescale. More broadly, the present work establishes a wave-packet-level framework for CTTS formation from core excitation in solution and provides a basis for connecting present CHC-type measurements with future direct time-resolved experiments.

\section{Computational Methods}
Our TDCI method is based on an electronic wavefunction $\Psi(t)$, represented as a linear combination of Slater determinants $\Phi_{I}$, with time-dependent expansion coefficients $C_{I}(t)$:
\begin{equation}
\label{eq1tdci}
\Psi(t) = \sum_{I} C_{I}(t) \Phi_{I}\,.
\end{equation}
The field-free Hamiltonian and the initial wave function are obtained by performing the time-independent calculation for the ground state with an \textit{ab initio} CI method of choice. Next, we numerically solve the time-dependent Schr{\"o}dinger equation:
\begin{equation}
\label{eq2tdci}
i\dot{\bm{C}}(t) =\bm{H}(t)\bm{C}(t)\,,
\end{equation}
thereby evolving the CI coefficients in time with the fourth-order Runge-Kutta scheme (RK4). This integrator provides additional numerical stability through the inherent damping of highly oscillatory components, which are typical for core-excited determinants and might otherwise require an impractically short time step. The propagation does not require an explicit construction of individual excited state CI vectors since a direct $\bm{HC}$ product inherently includes all possible states in the simulation. Specific observables can subsequently be analyzed by projection onto selected stationary states, while more general quantities, such as the electronic density or fragment charges, are readily available during the propagation. The amount of electron correlation is determined by the underlying CI ansatz. The configuration interaction singles (CIS) is the simplest and most computationally efficient option, restricting the wavefunction to singly-excited determinants. A more advanced option is the multireference configuration interaction (MRCIS), which relies on an active space selection, considering all determinants within and single-excited references built upon them. Multi-reference treatment is needed in cases when orbital relaxation is important; an effect which is absent in the CIS formalism. Complete active space CI (CASCI) is also a valid choice. To further reduce the computational costs, selected occupied orbitals can be frozen while all excitations of the $1s$ orbital of the aluminum ion are retained. 

We opted for an exclusive $1s$ CIS variant based on Hartree-Fock molecular orbitals for most of the calculations. This allows us to model the hydrated ion with two solvation layers, providing meaningful insight on how CTTS states form. Alternative methodological choices are discussed in the relevant sections.

To include the laser field, we add an extra term to the field-free Hamiltonian (${H}_0$):
\begin{equation}
\label{eq3field}
\bm{H}(t) = \bm{H}_0 - \bm{\mu}\cdot \bm{d}\,E(t)\,,
\end{equation}
which contains the transition-dipole matrix $\bm{\mu}$, the unit vector along the field polarization direction $\bm{d}$, and the time-dependent electric field amplitude $E(t)$. This allows for different pulse shapes such as the $\delta$-pulse, continuous-wave, or the transform-limited pulse. 

We employ standard atomic basis sets for all calculations, using the cc-pVDZ set for the ground state calculations and more diffuse def2-TZVPD on the aluminum atom ($6s5p3d$) for the time-dependent simulations. This choice allows for a more accurate description of the outgoing electron. However, in practice the outgoing electron is low in energy and moves slowly away from the ion. Adding more diffuse basis functions does not efficiently extend the spatial region relevant to the dynamics. While uncontracting the basis would further improve the energetics, it leads to prohibitively expensive and numerically unstable simulations.

The TDCI method is implemented in a development version of the GPU-accelerated code \mbox{TeraChem}.\cite{terachem21,Ufimtsev2008,Ufimtsev2009a,Ufimtsev2009b,Fales2015,tdci,durden2022} We also employ the OpenMolcas software for time-independent calculations.\cite{manni2023openmolcas}

\section{Results and Discussion}
 To explore the CTTS process, we consider a hierarchy of model systems: the isolated Al$^{3+}$ ion, the simplest ``charge-transfer dimer'' Al(H$_2$O)$^{3+}$, the compact first-shell complex Al(H$_2$O)$_6^{3+}$, and an extended Al(H$_2$O)$_{18}^{3+}$ cluster that mimics two solvation shells and includes more loosely bound water molecules. For each system, we can propagate the electronic wavefunction under an explicit X-ray field. Short, transform-limited (TL) pulses (bandwidth $1$~eV) are used to mimic the sudden preparation of 1$s$-excited wave packets across the near-edge absorption band. 
By monitoring partial charges on the ion and water molecules as well as the real-space electron density, we follow the migration of the excited electron from metal-centered $\sigma^\ast$ orbitals into more solvent-localized states, and we show how this migration accelerates as the excitation energy is tuned deeper into the CTTS band. Continuous-wave (CW) driving is also used to explore the relative rate of the outgoing electronic wave on short timescales, before the size limits of the model become notable. These simulations also show how the increasing density of solvent states drives the transition from discrete, Rabi-like dynamics observed in an isolated ion to quasi-irreversible charge transfer via rapid delocalization into a dense manifold of solvent-supported states.


\subsection{Electronic structure: From atoms to bulk}
To characterize the hydrated aluminum ion, we performed a detailed analysis of molecular orbitals, excited state energies, and characters using various quantum chemistry methods. Here we summarize the main results, with all details provided in the Electronic supplementary material (ESI), so as to keep the focus on the charge transfer process.

The Al$^{3+}$ Hartree-Fock (HF) calculation naturally yields atomic-like molecular orbitals (MOs) of an $s,p,$ and $d$ character. The first bright excitation corresponds to the $1s\rightarrow 3p$ transition from the $1s^22s^22p^6$ ground state configuration, in agreement with the $\Delta l=\pm 1$ selection rule for atomic transitions, where $l$ is the angular momentum quantum number. The simple CIS/cc-pVDZ method yields the transition energy of 1592.3~eV while full CI (FCI) gives 1587.9~eV. We further quantified the single-excitation character using $\omega=\mathrm{Tr}(\gamma\gamma^T)$, where $\gamma$ is the one-particle density matrix. It is unity for the CIS method (no higher determinants are available by design), and $\omega = 0.97$ for the FCI calculation. This indicates that the transition is dominated by single-excitation character, and electron correlation effects are minor. This is a crucial result that still holds even for the hydrated ions and supports the use of the computationally effective CIS method.

\begin{figure}[h!]
\centering
\includegraphics[width=.4\textwidth]{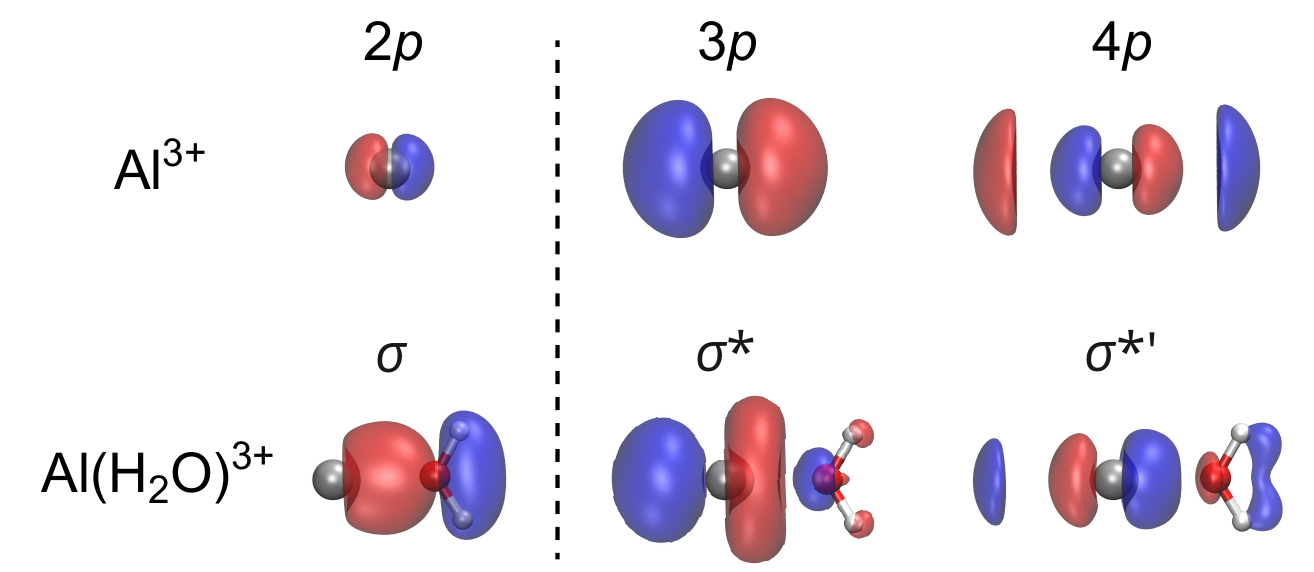}
\caption{Hartree-Fock orbitals of $p$-character for Al$^{3+}$ (upper row) transformed to selected $\sigma$ orbitals in the Al(H$_2$O)$^{3+}$ case (lower row).}%
\label{fig:al_alwat_orbs}
\end{figure}

The addition of water molecules to the aluminum ion transforms the initial unoccupied atomic-like $3p$ MOs into $\sigma$-like orbitals oriented along the aluminum-oxygen axes, concentrating the electron density in the region between the atoms. In the literature, such low-lying diffuse core-excited states are often discussed as having frustrated Rydberg-like character.\cite{carter2022choice} Here, however, we stay with the $\sigma$/$\sigma^*$
 labeling as it more directly reflects their orbital shape and Al–O antibonding character. The higher-energy bright state (originally $4p$) has a similar $\sigma$ shape, but with more electron density shifted towards the coordinating water and increased antibonding character (Fig. \ref{fig:al_alwat_orbs}). The solvated Al(H$_2$O)$_{6}^{3+}$ system has the excitation energy of 1590.9~eV (CIS/cc-pVDZ) for the lowest-energy bright transition, compared to the experimental spectral onset of 1565~eV,\cite{muchova_coreholeclock_ctts_2024} which is a typical shift for core-excited state calculations that are not specifically tailored.\cite{besley2009time} A closer agreement with the experimental value would require higher electron correlation and drastically higher computational costs. Full characterization of 1$s\xrightarrow{}3p$ transition with CIS to FCI accuracy can be found in ESI.


\subsection*{Characterization of target system: Al(H$_2$O)$_{18}^{3+}$}

\begin{figure}[h!]
\centering
\includegraphics[width=.48\textwidth]{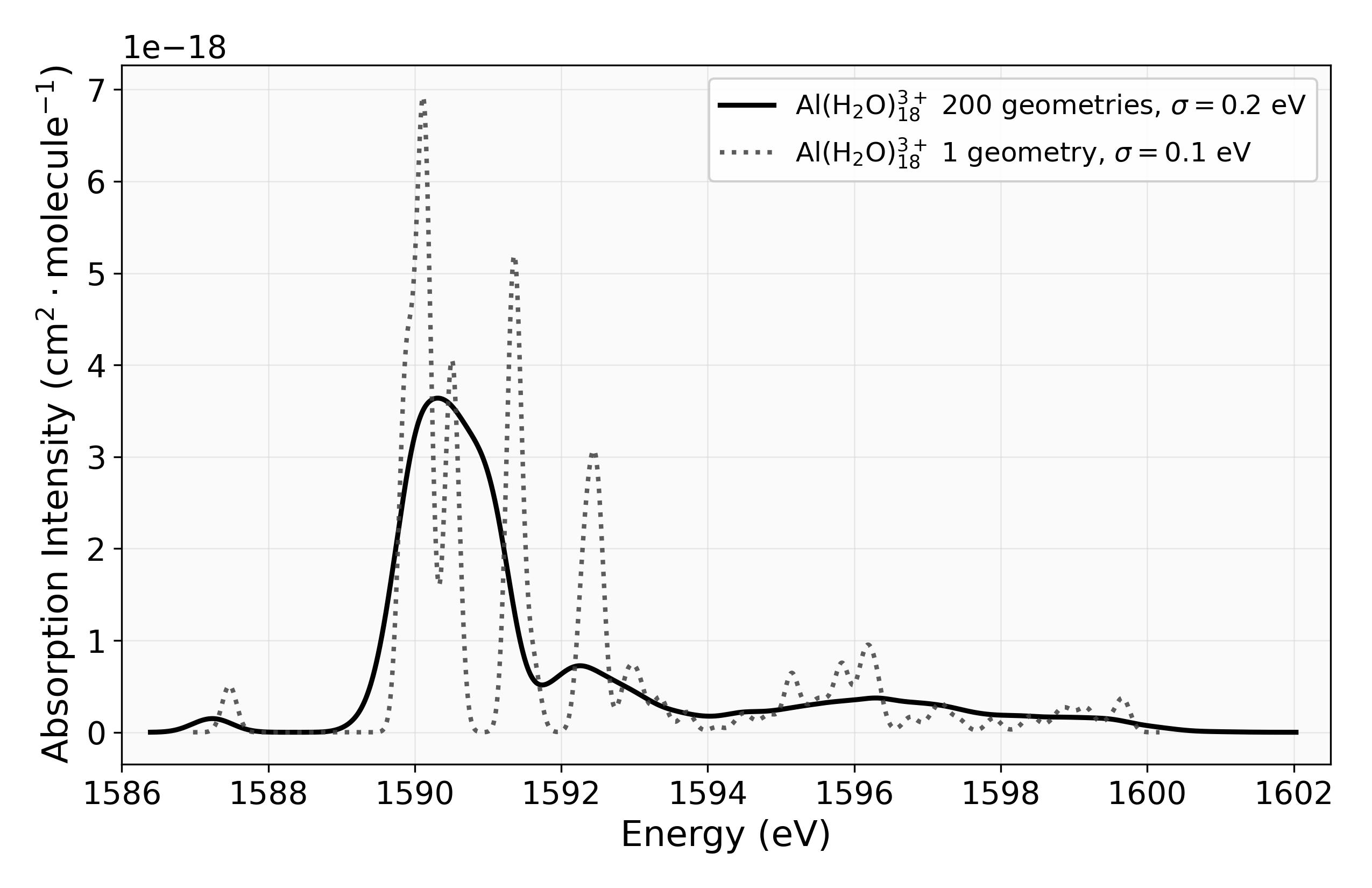}
\caption{Calculated absorption spectrum of the core $1s$ excited states of Al(H$_2$O)$_{18}^{3+}$ obtained with the CIS method with def2-TZVPD(Al), cc-pVDZ(O,H) basis. The first 50 electronic transitions were broadened with a Gaussian function of a given standard deviation $\sigma$.}%
\label{fig:alwat18_spec}
\end{figure}

In the time-dependent simulations, we focus exclusively on the Al(H$_2$O)$_{18}^{3+}$ aqua complex. Because the real liquid is characterized by a statistical distribution of geometries, Fig.~\ref{fig:alwat18_spec} shows the CIS spectrum obtained from 200 geometries sampled from classical molecular dynamics simulations.\cite{muchova_coreholeclock_ctts_2024} For the time-dependent simulations, we selected a single representative geometry from the ensemble. The spectrum evaluated for this geometry is also shown in Fig.~\ref{fig:alwat18_spec}. The spectra were obtained within the reflection principle/nuclear ensemble method.\cite{srsen2020limits} Compared to the smaller hexaaqua complex, Al(H$_2$O)$_{18}^{3+}$ shows a substantially increased number of available excited states. In general, the lowest-energy peak at $\sim$1590--1592.5~eV corresponds to the transitions to the $\sigma^*$ orbitals (akin to atomic $3p$ states), second sub-peak at 1592.5--1595 eV corresponds to the $\sigma^*$ transitions with higher $4p$ character and the band around 1595--1600 eV corresponds primarily to the excitations to the second solvation shell orbitals on water molecules. All transitions are expected to lie below the ionization threshold. The Koopmans-like ionization energy (i.e. 1s orbital energy) estimated at the same level of theory is $\sim$1600 eV.   

\subsection{Interaction with a finite transform-limited pulse}

\begin{figure}[h!]
\centering
\includegraphics[width=.5\textwidth]{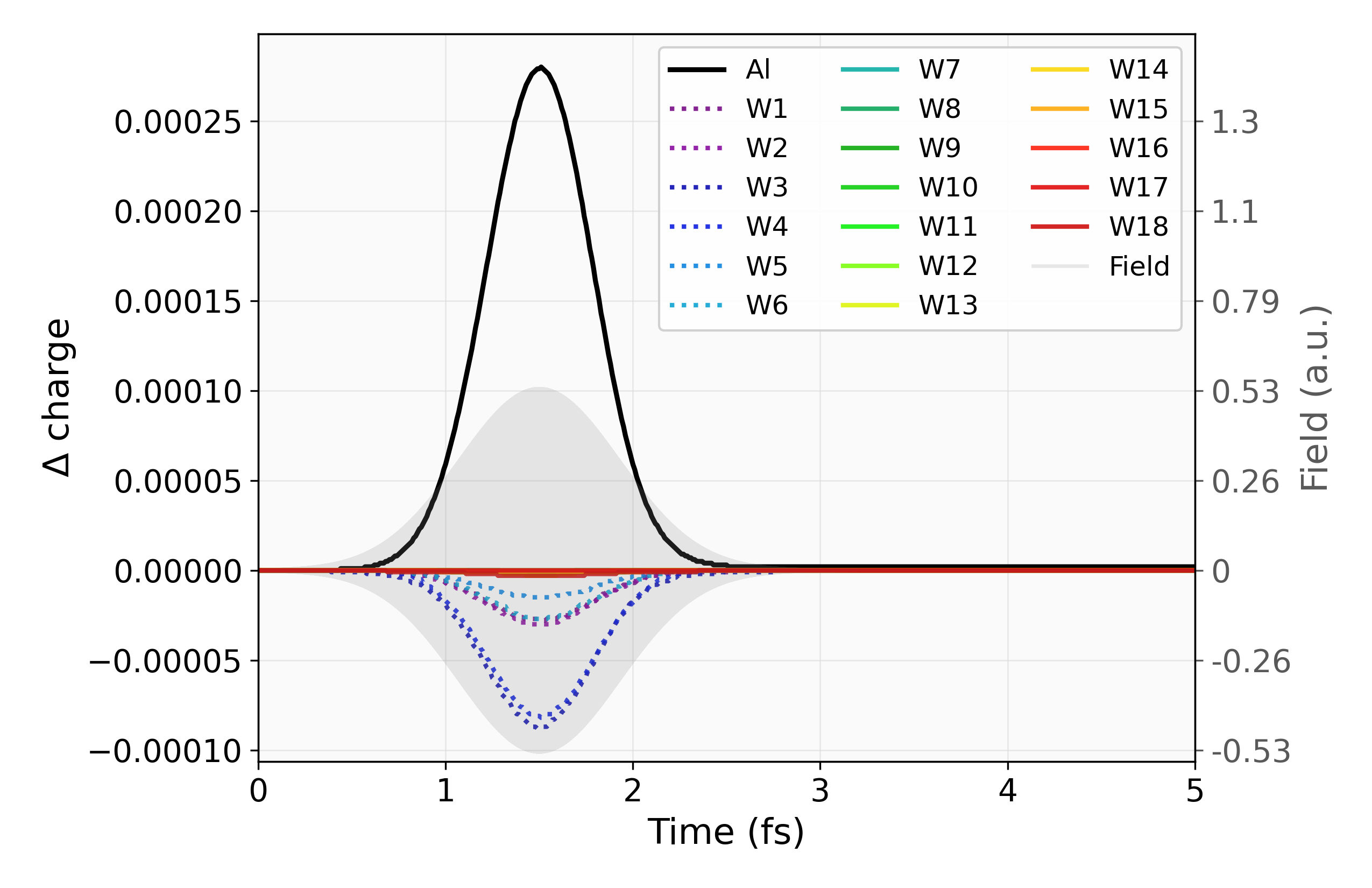}
\caption{Charge difference from the initial partial charges in the Al(H$_2$O)$_{18}^{3+}$ system during a TD-CIS simulation with 1~fs TL pulse centered at 1.5~fs at energy of 1585~eV. Al (black), first solvation shell water molecules (dashed lines), second solvation shell water molecules (full lines).}%
\label{fig:chg1585}
\end{figure}

First, we simulate the CT process in the Al(H$_2$O)$_{18}^{3+}$ using the TD-CIS method with the TL pulse (FWHM = 1~fs, centered at $t$=1.5~fs, peak intensity 1$^{20}$ W/m$^2$, $x$ polarization direction). We performed a series of simulations with pulse energies scanning across the absorption band (1585--1605~eV). To reveal the underlying electronic dynamics, we analyze the differences in the partial charges of Al, water molecules in the first solvation shell (W1-W6), and those in the second solvation shell (W7-W18). The charges are calculated using a one-particle density matrix with Voronoi formalism. Fig.~\ref{fig:chg1585} displays the results for the pulse energy of 1585~eV, targeting the region below the first absorption peak. Due to the uncertainty in energy, the pulse off-resonantly perturbs the ground state, causing a partial excitation of the electron to the first solvation shell. The excitation subsides once the pulse has ended. The absolute differences in the TD simulations can be very low to remain in the linear regime and should therefore be interpreted in relative rather than absolute terms.

\begin{figure}[h!]
\centering
\includegraphics[width=.75\textwidth]{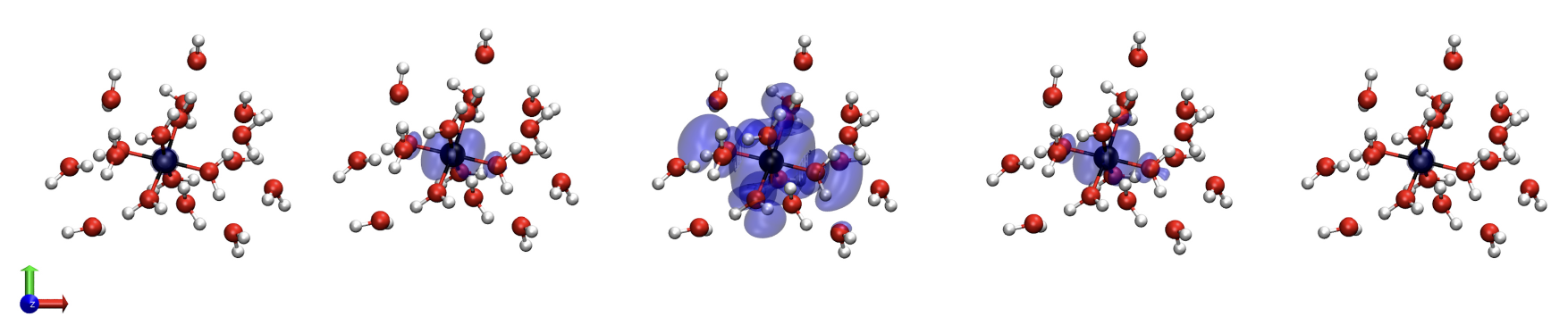}
\caption{Electron density difference in the Al(H$_2$O)$_{18}^{3+}$ system under the 1~fs TL pulse centered at 1.5~fs at energy of 1585~eV. The five time profiles are taken at 0, 0.75, 1.5, 2.25, and 3.0~fs.}%
\label{fig:dens1}
\end{figure}

To visualize the charge transfer process, we show the electron density difference between time $t$ and $t_0$ in Fig.~\ref{fig:dens1}. The shape of the excited density resembles a $p$-wave in agreement with the selection rules. The overall intensity of the process is very low, compared to the following on-resonance simulations. 

\begin{figure*}[t]
\centering
\includegraphics[width=1.0\textwidth]{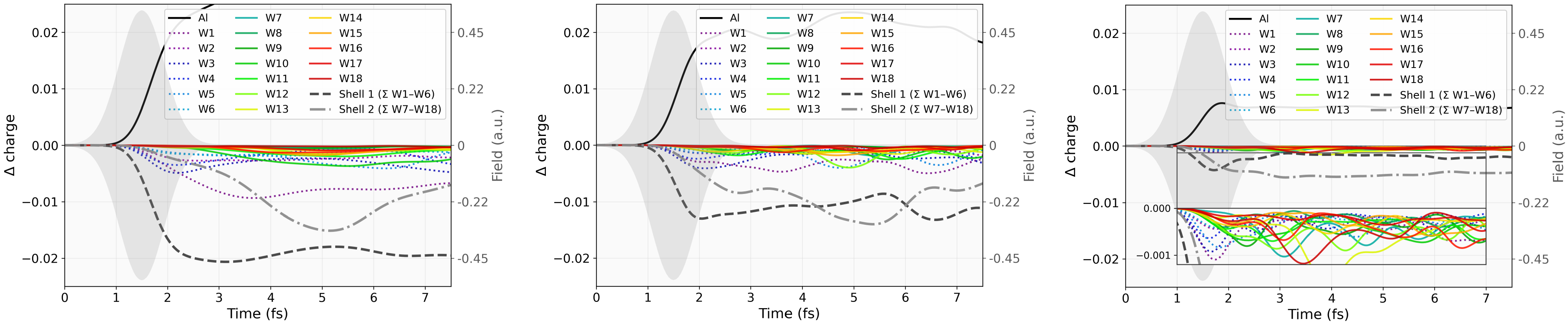}
\caption{Difference from the initial partial charges in the Al(H$_2$O)$_{18}^{3+}$ system during a TD-CIS simulation  with the 1~fs TL pulse centered at 1.5~fs for pulse energies of 1590~eV (left), 1592.5~eV (middle), and 1595~eV (right).}%
\label{fig:chg3}
\end{figure*}

Fig.~\ref{fig:chg3} shows the evolution of the partial charges for laser pulse energies of 1590~eV, 1592.5~eV, and 1595~eV across the main absorption peak. The pulse induces excitations that persist after the field is switched off (around 3~fs). The overall transition intensity decreases with increasing pulse energy as expected from the spectrum. At the same time, the spatial characteristics change, the lower-energy pulse transfers the charge preferentially to the first solvation shell whereas the higher-energy pulse favors transfer to the second solvation shell. This is consistent with the nature of the electronic transitions discussed above in the analysis of the absorption spectrum peaks. The pulse in all cases induces subsequent charge dynamics arising from the existing coherence between the excited states. Within the CIS framework, the populations of these states remain unchanged without field being present, but their relative phases continue to evolve. This charge dynamics is experimentally observable. The higher the pulse energy, the more rapid the charge migration, which is due to the larger number of contributing states and their broader energy spread. This points to an interesting dichotomy, where a stationary coherent superposition of eigenstates gives rise to dynamically evolving charge migration in time.

\begin{figure}[h!]
\centering
\includegraphics[width=.4\textwidth]{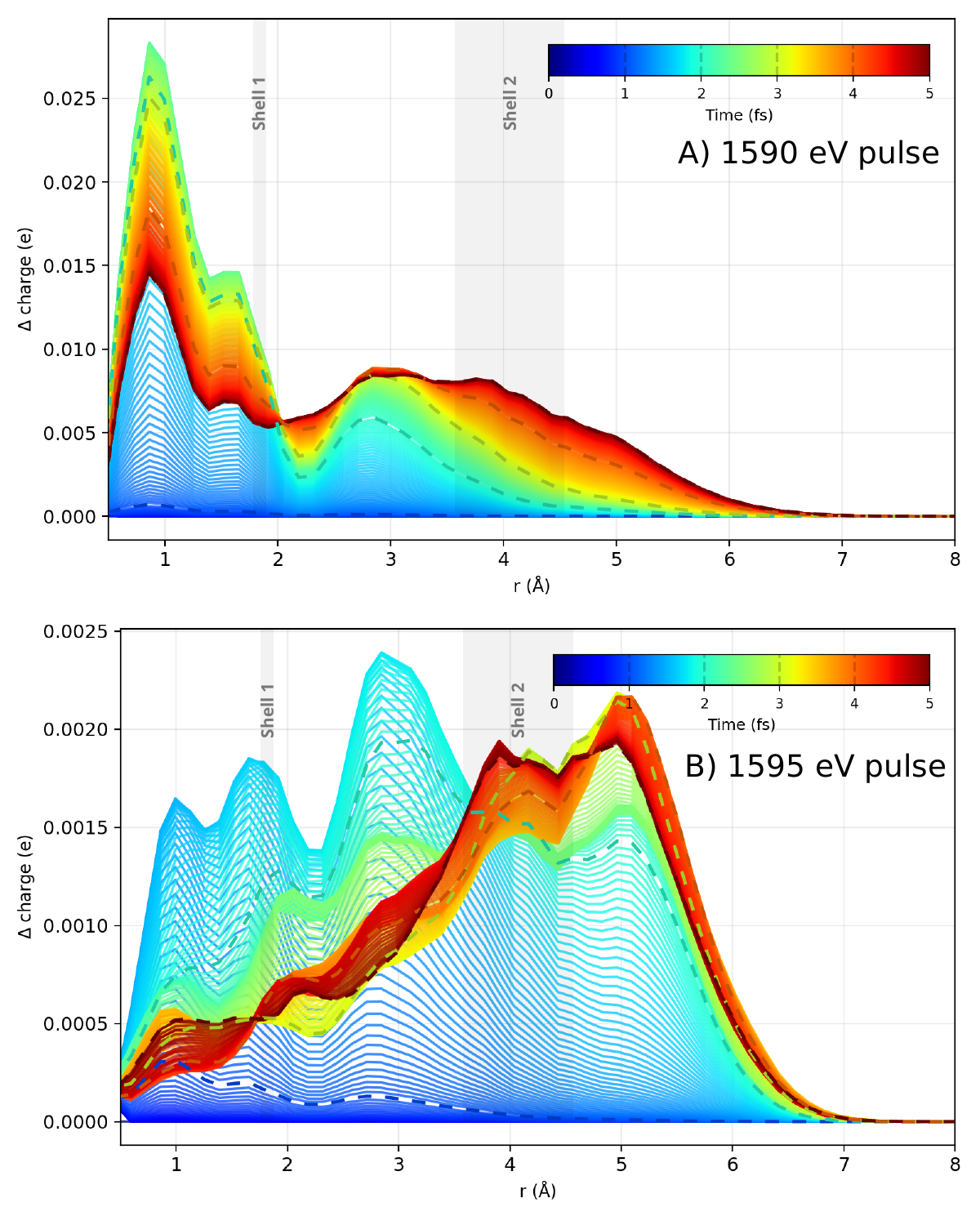}
\caption{Radial profile of the electron density difference for Al(H$_2$O)$_{18}^{3+}$ system under TL pulse excitation at 1590~eV (A, upper) and 1595~eV (B, lower). Line color changes from blue (0~fs) to red (5~fs) with 0.02~fs spacing and dashed lines at each whole fs.}%
\label{fig:radial}
\end{figure}

Moreover, it implies that the charge delocalization is not a sequential two-step process. From the beginning, the pulse concurrently populates states with electron density distributed in both the first and second solvation shells, only with different rates. To visualize this aspect more clearly, Figure~\ref{fig:radial} shows calculated radial profiles of the excited electron obtained by integration of electron density difference in the solvation shells around the Al ion. We can observe that the electron is initially excited in the vicinity of the Al atom, and at later times, the density is shifted outwards towards the surrounding water molecules. This apparent motion is again obtained even though there is no population transfer between states after the pulse has ended (3~fs mark, yellow dashed line). 

It could thus appear as if the excited electron density migrates outward more rapidly at higher pulse energies. For the lower-energy pulse (1590~eV), the density is concentrated in the area between the Al atom and the first solvation shell ($0-1.8$ \si{\angstrom}) and in the area between the first and second solvation shells ($\sim 2-4$ \si{\angstrom}). For the higher-energy pulse (1595~eV), the final density resides mostly on the second solvation shell (5 \si{\angstrom}). If the simulation were extended beyond 5~fs, we would observe a partial reflection of the electron density from the outermost region and further oscillations. This reflects the finite spatial extent of the basis-set description, which effectively confines the electron within the simulated cluster (a particle in a box scenario). In a larger system with a third solvation shell or simply in the presence of a continuum, the electron would have more pathways for further delocalization. However, any subsequent observable dynamics would still be limited by the core-hole lifetime, and hence we restrict the discussion to the presented time window. Further increasing the excitation energies diminishes the overall intensity of excitations in agreement with the experimental observations. 

The experimental Auger spectrum provides an indirect probe of the evolving core-excited charge-transfer dynamics. To connect the quantum-mechanical simulation to experiment, the following picture is useful. Because the core-excited state has a finite lifetime (around 1.5~fs for Al), Auger decay samples the evolving electronic configuration at different stages of the wavepacket dynamics. Based on how far the excited electron escaped, we observe a different screening of the core charge during the Auger process. Collapse of the wavefunction into the manifold of final states leads to different electronic configurations, each associated with a distinct Auger electron kinetic energy and thus contributing differently to the spectrum. Explicit access to these final states in the simulation would therefore be required to model the Auger decay itself. However, reproducing the experimental spectrum would additionally require averaging over a statistical ensemble of molecular geometries and accounting for nuclear motion, since the measured signal represents an effective time- and ensemble-averaged sum over all such processes.

\subsection{Interaction with continuous-wave pulse}
\begin{figure}[h!]
\centering
\includegraphics[width=.45\textwidth]{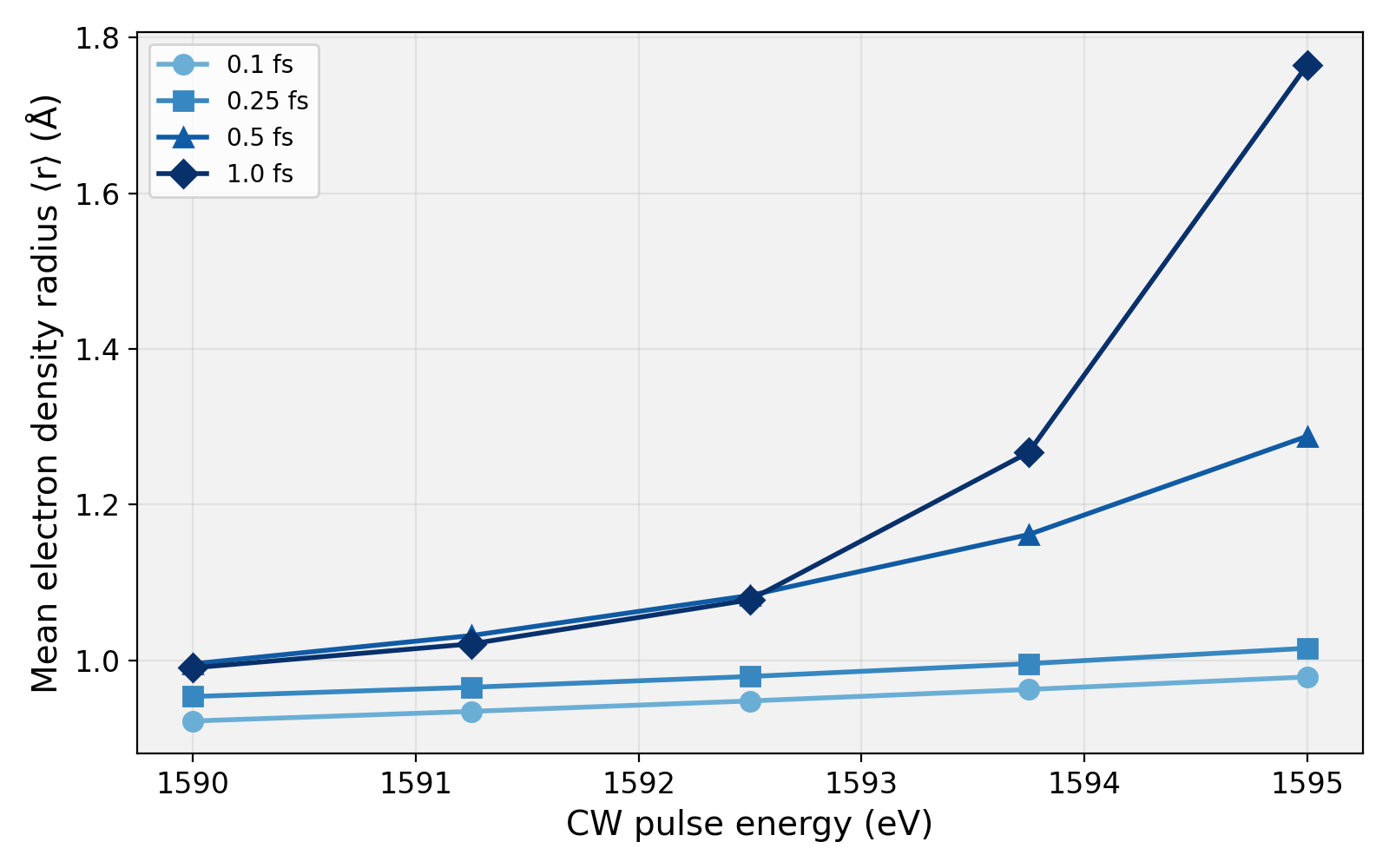}
\caption{Mean radial position calculated from the electron-density-difference in the Al(H$_2$O)$_{18}^{3+}$ system with CW pulse of given energies at various times.}%
\label{fig:radialCW}
\end{figure}

Finally, we simulate the system with a continuous wave pulse to mimic the narrow-band, energy-selective excitation limit. For the TL pulsed simulations, we analyzed solvation shell-resolved charge differences in order to visualize how the excited electron density is redistributed among the ion, the first solvation shell, and the second solvation shell in real time. For the CW simulations, however, we adopt a different observable. Here the goal is not to resolve transient charge flow between predefined shells, but rather to extract a simple scalar measure of how far the excited electron extends from the ion as a function of excitation energy. This choice is motivated by the core-hole-clock picture, in which charge transfer is often interpreted in terms of the rate at which the excited electron delocalizes away from the initially localized core-excited state. We therefore analyze the mean radial position of the excited density, which provides a compact proxy for the apparent escape of the electron from the ion.

Fig.~\ref{fig:radialCW} shows the radial extent of the electron in the time interval from 100~as to 1.0~fs of the simulation. The figure clearly shows the dependence of the charge delocalization rate on the pulse energy. 
The extracted radial dynamics provides merely a simplified classical view of the early-time charge delocalization. The propagated wavepacket represents a coherent superposition of states with finite lifetimes, which would interfere when coupled to the final states. Moreover, the CW field, which is closest to what the molecule experiences in the synchrotron or X-ray free-electron laser (XFEL) experiments, is constantly pumping additional density to the already evolving wavefunction. Although this view represents a substantial simplification of the full quantum dynamics, it still reveals a clear exponential trend. 

\subsection{From Rabi oscillations to quasi-irreversible dynamics}
\begin{figure*}[t]
\centering
\includegraphics[width=1.0\textwidth]{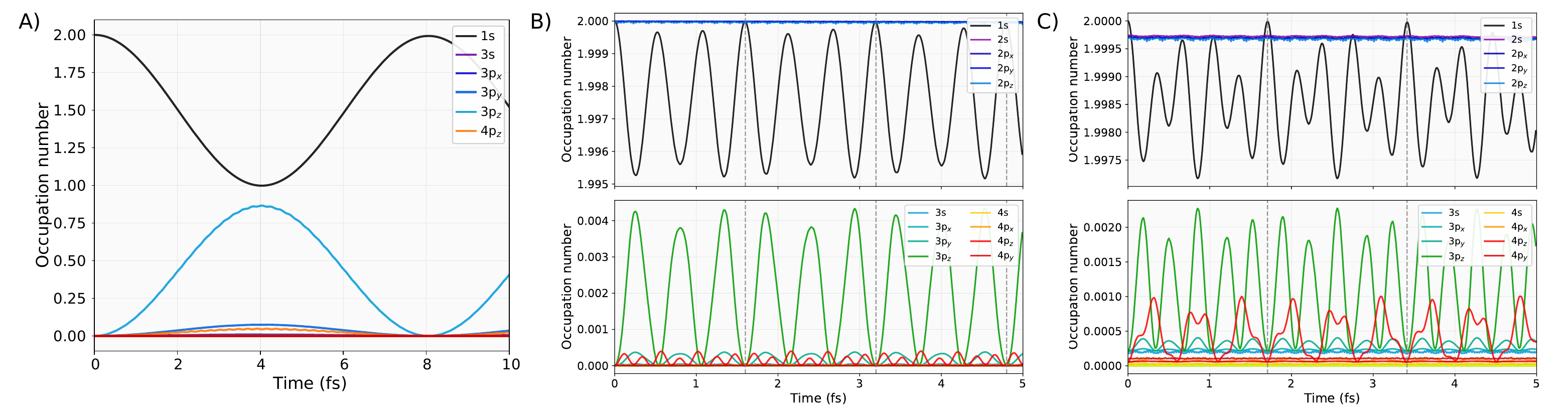}
\caption{Molecular orbital occupation numbers from one electron reduced density matrix calculation of the Al$^{3+}$ system during a) TD-CIS/cc-pVDZ simulation with a CW laser tuned to 1$s\rightarrow 3p$ transition b) tuned to 1600~eV energy and c) TD-MRCIS(10,9) simulation with a CW laser tuned to 1600~eV.}%
\label{fig:rabi}
\end{figure*}

In this last section, we illustrate how the observed CTTS dynamics can be rationalized in terms of simple field-driven oscillatory dynamics between two states. We begin with a TD-CIS/cc-pVDZ dynamics for the isolated Al$^{3+}$ ion, considering all the available orbitals and tuning the pulse exactly to the 1$s\rightarrow3p$ excitation energy (1592.26~eV). Fig.~\ref{fig:rabi}A shows how orbital occupation numbers from the one-electron density matrix calculation evolve in time. As expected, the electron is transferred from the 1$s$ orbital into the initially unoccupied 3$p$ and 4$p$ orbitals (laser axis is not aligned with any specific direction). After reaching the maximum, the electron returns back to the initial state, completing a Rabi cycle with a period of 8~fs. 

This situation corresponds to exact resonance with the target state. In more realistic cases, however, many bright states are available, and the laser field is off-resonant, which drives multiple concurrent excitations. The overall strength depends both on the detuning of the state energies $\omega$ to the laser central frequency $\omega_0$ and on their couplings to the ground state $V_{ab}$ (e.g., transition dipole between ground state $a$ and final state $b$). The total Rabi frequency for each transition is then:

\begin{equation}
\label{eq1det}
\tilde{\Omega}_{ab} \equiv \sqrt{(\omega-\omega_0)^2+\frac{|V_{ab}|^2}{\hbar^2}}
\,,
\end{equation}
and the peak amplitude $P_b^{\max}$ for each transition is then reduced with increased detuning $\Delta$ of the laser:
\begin{equation}
\label{eq2det}
P_{ab}^{\max}=\frac{|V_{ab}|^2/\hbar^2}{(\omega-\omega_0)^2+|V_{ab}|^2/\hbar^2}
=\frac{\Omega_R^2}{\Delta^2+\Omega_R^2},
\quad \Omega_R\equiv \frac{|V_{ab}|}{\hbar}.
\end{equation}

An example of this behavior is presented in Fig.~\ref{fig:rabi}B for an otherwise identical TD-CIS simulation setup with a laser tuned to 1600~eV. Both the 3$p$ and 4$p$ MOs are populated, but the following dynamics is more intricate since it involves an interplay of two dominant transitions. The overall Rabi period, indicating return to the initial state, decreases to 1.6~fs, and the overall amplitude becomes significantly lower. 



Lastly, we discuss the MRCIS(10,9) dynamics, which includes a larger determinant space and both a denser manifold of electronic states and additional electron-correlation effects. The corresponding dynamics shown in Fig.~\ref{fig:rabi}C is even more complex than in the previous case. Nevertheless, the system can still be driven back close to the initial state, with a period of about 1.7~fs. In these examples, we discussed a single ion with a relatively small atomic basis offering a limited set of accessible states and excited state determinants. Increasing any of these parameters will inevitably lead to a more complicated dynamics, extending the overall return period to the initial state indefinitely. Indeed, a simple addition of water molecules provides reversible dynamics only in a highly truncated CIS space. The quantum-mechanical picture here is that the dynamics stems from many interacting states and can be ultimately driven back. After all, the total wavefunction remains pure and evolves unitarily. Therefore, CTTS-like behavior appears effectively irreversible because population disperses over a dense manifold of states with many interfering frequencies and extremely long recurrence times.

\section*{Conclusions}


In this work, we introduced a fully time-dependent, wavefunction-based quantum-chemistry framework for following K-edge core excitation in solution and for real-time observation of CTTS-state formation. By explicitly propagating the electronic wavefunction in an applied X-ray field, we include the full manifold of electronically accessible configurations and directly visualize how a core-excited wave packet evolves in space and time. For Al$^{3+}$ clusters, ranging from the bare cation to a system with two solvation shells, the simulations show that a well-defined initial $p$-like wave packet is created through direct population of many states, and that subsequent charge migration is driven by coherences established during the excitation step rather than by a simple sequential "excite and then transfer" mechanism. A central conceptual outcome is therefore that CTTS formation from core levels in solution is intrinsically a many-state problem. In the isolated ion and in very small basis-set models, the dynamics can still resemble few-state Rabi oscillations, with clear recurrences of the initial state. As the density of solvent-supported states increases in larger clusters, however, the dynamics changes from Rabi-oscillation-like behavior to an apparently irreversible delocalization, as the population is dispersed over many frequencies. This provides a physically transparent explanation for why Rabi oscillations are clear in minimal models, whereas larger clusters exhibit delocalization-like transfer that appears effectively irreversible.

Placing these findings into the context of core-hole-clock (CHC) spectroscopy, our results suggest that CHC "time" extracted from intensity ratios is not a universal observable in this regime, but depends on how the core-excited manifold is populated and on the degree of coherence between localized and delocalized channels. In our real-time simulations, the X-ray field coherently excites multiple intermediate states with varying excitonic and CTTS character, and under continuous-wave excitation this evolving superposition is continuously driven. The quantity accessed through the Auger observables should therefore be viewed as an effective dynamical response of the driven core-excited manifold, rather than as a unique post-excitation charge-transfer time.

Methodologically, the present results highlight the usefulness of real-time wavefunction-based propagation for core-hole excitations, where dense excited-state manifolds, strong coherence, and ultrafast driving fields make a purely stationary picture difficult to interpret and where approximate kernels, self-interaction, and pseudopotentials can limit the reliability of real-time TDDFT.

Several important ingredients remain for future work. Most notably, Auger decay is not included explicitly here; incorporating it either through scattering-state continua or through effective non-Hermitian or Liouville--von Neumann/Lindblad treatments would enable a more direct connection to observable spectra and to core-hole-clock analyses on the same footing. Improved treatment of the outgoing electron, for example through complex absorption potentials (CAP) or more flexible continuum-oriented basis representations, would further reduce boundary artifacts and extend the accessible dynamical window.\cite{krause2014strong,marante2014hybrid} In addition, solvent polarization and continuum effects could refine energetics and spectral alignment, although the primary mechanistic conclusions are expected to remain robust.

Looking ahead, one can envision moving from "internal clocks" to direct pump--probe timing of core-excited charge transfer in solution. Attosecond and sub-femtosecond X-ray pulse pairs and attosecond X-ray spectroscopy in liquids have recently been demonstrated,\cite{guo2024experimental} indicating that time-resolved measurements of core-excited wave-packet formation are within reach. A particularly attractive direction would be to combine a core-edge pump, which launches the CTTS wave packet, with a delayed X-ray probe that reads out the evolving localization and delocalization either spectroscopically, through time-resolved X-ray absorption or emission, or through channel-resolved resonant Auger yields. Our time-dependent framework provides clear predictions for the early-time buildup of CTTS character that such experiments could directly test.

\section*{Acknowledgements}
This work was supported by a grant award DE-SC0021643 from the Computational and Theoretical Chemistry program of the Office of Basic Energy Sciences, U.S. Department of Energy.  J.S. acknowledges a postdoctoral fellowship from the Institute for Advanced Computational Science (IACS). E.M. acknowledges the support of the Czech Science Foundation project number 25-15408S. P.S. acknowledges the support by the Czech Science Foundation project number 26-22810S. This work was supported by the project ``The Energy Conversion and Storage", funded as project No. CZ.02.01.01/00/22\textunderscore 008/0004617 by Programme Johannes Amos Comenius, call Excellent Research.

\section*{Data Availability Statement}
The data underlying this study are openly available in Zenodo at \href{https://doi.org/10.5281/zenodo.21833445}{10.5281/zenodo.21833445}.

\section*{Supporting information}
The following files are available free of charge.
\begin{itemize}
  \item ESI.pdf: Additional computational results.
\end{itemize}

\section*{Table of Contents Graphic}
\begin{figure}[h]
  \centering
  \includegraphics[width=3.25in,height=1.75in]{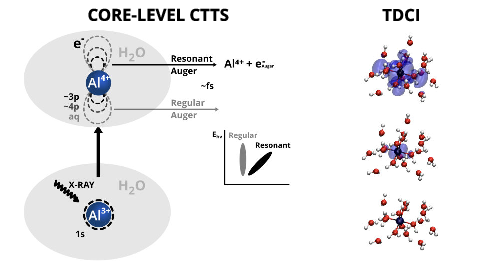}
  \caption{For Table of Contents Only}
\end{figure}



\printbibliography

@article{tdci,
author = {Peng, Wei-Tao and Fales, B. Scott and Levine, Benjamin G.},
title = {Simulating Electron Dynamics of Complex Molecules with Time-Dependent Complete Active Space Configuration Interaction},
journal = {Journal of Chemical Theory and Computation},
volume = {14},
number = {8},
pages = {4129-4138},
year = {2018},
doi = {10.1021/acs.jctc.8b00381},
    note ={PMID: 29986143},
URL = { 
            https://doi.org/10.1021/acs.jctc.8b00381 
},
eprint = { 
        https://doi.org/10.1021/acs.jctc.8b00381
}
}

@article{herbert_hydrated_electron_2017,
  author  = {Herbert, John M.},
  title   = {The Hydrated Electron},
  journal = {Annual Review of Physical Chemistry},
  year    = {2017},
  volume  = {68},
  number  = {1},
  pages   = {447-472},
  doi     = {10.1146/annurev-physchem-052516-050816},
  URL     = {https://doi.org/10.1146/annurev-physchem-052516-050816},
  eprint  = {https://doi.org/10.1146/annurev-physchem-052516-050816}
}

@article{carterfenk_birth_hydrated_electron_2023,
  author  = {Carter-Fenk, K. and Johnson, C. J. and Herbert, John M.},
  title   = {Birth of the Hydrated Electron via Charge-Transfer-to-Solvent Excitation of Aqueous Iodide},
  journal = {The Journal of Physical Chemistry Letters},
  year    = {2023},
  volume  = {14},
  pages   = {870--878},
  doi     = {10.1021/acs.jpclett.2c03460}
}

@article{rabinowitch_electron_transfer_spectra_1942,
  author  = {Rabinowitch, Eugene},
  title   = {Electron Transfer Spectra and Their Photochemical Effects},
  journal = {Reviews of Modern Physics},
  year    = {1942},
  volume  = {14},
  number  = {2-3},
  pages   = {112--131},
  doi     = {10.1103/RevModPhys.14.112},
  URL     = {https://doi.org/10.1103/RevModPhys.14.112},
  eprint  = {https://doi.org/10.1103/RevModPhys.14.112}
}

@article{messina_real_time_ctts_2013,
  author  = {Messina, Fabrizio and Br{\"a}m, Olivier and Cannizzo, Andrea and Chergui, Majed},
  title   = {Real-Time Observation of the Charge Transfer to Solvent Dynamics},
  journal = {Nature Communications},
  year    = {2013},
  volume  = {4},
  number  = {1},
  pages   = {2119},
  doi     = {10.1038/ncomms3119},
  URL     = {https://doi.org/10.1038/ncomms3119},
  eprint  = {https://doi.org/10.1038/ncomms3119}
}

@article{barthel_sodide_ctts_2003,
  author  = {Barthel, E. R. and Martini, I. B. and Keszei, E. and Schwartz, B. J.},
  title   = {Solvent Effects on the Ultrafast Dynamics and Spectroscopy of the Charge-Transfer-to-Solvent (CTTS) Reaction of Sodide},
  journal = {The Journal of Chemical Physics},
  year    = {2003},
  volume  = {118},
  pages   = {5916--5931}
}

@article{okuyama_ctts_iodide_trpes_2016,
  author  = {Okuyama, Haruki and Suzuki, Yoshi-Ichi and Karashima, Shutaro and Suzuki, Toshinori},
  title   = {Charge-Transfer-to-Solvent Reactions from I\textsuperscript{--} to Water, Methanol, and Ethanol Studied by Time-Resolved Photoelectron Spectroscopy of Liquids},
  journal = {The Journal of Chemical Physics},
  year    = {2016},
  volume  = {145},
  number  = {7},
  pages   = {074502},
  doi     = {10.1063/1.4960385},
  URL     = {https://doi.org/10.1063/1.4960385},
  eprint  = {https://doi.org/10.1063/1.4960385}
}

@article{kothe_ctts_dynamics_iodide_2015,
  author  = {Kothe, Alexander and Wilke, Martin and Moguilevski, Alexandre and Engel, Nicholas and Winter, Bernd and Kiyan, Igor Yu. and Aziz, Emad F.},
  title   = {Charge Transfer to Solvent Dynamics in Iodide Aqueous Solution Studied at Ionization Threshold},
  journal = {Physical Chemistry Chemical Physics},
  year    = {2015},
  volume  = {17},
  pages   = {1918--1925},
  doi     = {10.1039/C4CP02482F}
}

@article{lubcke_trpes_solvated_electron_2010,
  author  = {L{\"u}bcke, Andrea and Buchner, Franziska and Heine, Nadja and Hertel, Ingolf V. and Schultz, Thomas},
  title   = {Time-Resolved Photoelectron Spectroscopy of Solvated Electrons in Aqueous NaI Solution},
  journal = {Physical Chemistry Chemical Physics},
  year    = {2010},
  volume  = {12},
  number  = {43},
  pages   = {14629-14634},
  doi     = {10.1039/C0CP00847H},
  URL     = {https://doi.org/10.1039/C0CP00847H},
  eprint  = {https://doi.org/10.1039/C0CP00847H}
}

@article{kammrath_ctts_precursor_clusters_2005,
  author  = {Kammrath, A. and Verlet, J. R. R. and Bragg, A. E. and Cheshnovsky, O. and Neumark, D. M.},
  title   = {Dynamics of Charge-Transfer-to-Solvent Precursor States in \textit{I}\textsuperscript{--}(\ce{H2O})\_n Clusters},
  journal = {The Journal of Physical Chemistry A},
  year    = {2005},
  volume  = {109},
  pages   = {11475--11483}
}

@article{bruehwiler_core_spectroscopies_review_2002,
  author = {Br{\"u}hwiler, P. A. and Karis, O. and M{\aa}rtensson, N.},
  title = {Charge-Transfer Dynamics Studied Using Resonant Core Spectroscopies},
  journal = {Reviews of Modern Physics},
  year = {2002},
  volume = {74},
  number = {3},
  pages = {703--740},
  doi = {10.1103/RevModPhys.74.703},
  URL = {
    https://doi.org/10.1103/RevModPhys.74.703
  }
}

@article{muchova_coreholeclock_ctts_2024,
  author  = {Muchov{\'a}, Eva and Gopakumar, Gireesh and Unger, Isabel and {\"O}hrwall, Gunnar and C{\'e}olin, Denis and Trinter, Florian and Wilkinson, Ian and Chatzigeorgiou, Eleni and Slav{\'\i}{\v{c}}ek, Petr and Hergenhahn, Uwe and Winter, Bernd and Caleman, Carl and Bj{\"o}rneholm, Olle},
  title   = {Attosecond Formation of Charge-Transfer-to-Solvent States of Aqueous Ions Probed Using the Core-Hole-Clock Technique},
  journal = {Nature Communications},
  year    = {2024},
  volume  = {15},
  pages   = {8903},
  doi     = {10.1038/s41467-024-52740-5},
  URL     = {https://doi.org/10.1038/s41467-024-52740-5},
  eprint  = {https://doi.org/10.1038/s41467-024-52740-5}
}

@article{ottosson_core_hole_clock_review_2012,
  author  = {Ottosson, Niklas and {\"O}hrwall, Gunnar and Bj{\"o}rneholm, Olle},
  title   = {Ultrafast Charge Delocalization Dynamics in Aqueous Electrolytes: New Insights from Auger Electron Spectroscopy},
  journal = {Chemical Physics Letters},
  year    = {2012},
  volume  = {543},
  pages   = {1--11},
  doi     = {10.1016/j.cplett.2012.05.051}
}

@article{greenman_tdcis_strongfield_2010,
  author  = {Greenman, Loren and Ho, Phay J. and Pabst, Stefan and Kamarchik, Eugene and Mazziotti, David A. and Santra, Robin},
  title   = {Implementation of the Time-Dependent Configuration-Interaction Singles Method for Atomic Strong-Field Processes},
  journal = {Physical Review A},
  year    = {2010},
  volume  = {82},
  number  = {2},
  pages   = {023406},
  doi     = {10.1103/PhysRevA.82.023406}
}

@article{carlstrom_tdcis_molecular_2022,
  author  = {Carlstr{\"o}m, Stefanos and Spanner, Michael and Patchkovskii, Serguei},
  title   = {General Time-Dependent Configuration-Interaction Singles. I. Molecular Case},
  journal = {Physical Review A},
  year    = {2022},
  volume  = {106},
  number  = {4},
  pages   = {043104},
  doi     = {10.1103/PhysRevA.106.043104}
}

@article{carlstrom_tdcis_atomic_2022,
  author  = {Carlstr{\"o}m, Stefanos and Bertolino, Mattias and Dahlstr{\"o}m, Jan Marcus and Patchkovskii, Serguei},
  title   = {General Time-Dependent Configuration-Interaction Singles. II. Atomic Case},
  journal = {Physical Review A},
  year    = {2022},
  volume  = {106},
  number  = {4},
  pages   = {042806},
  doi     = {10.1103/PhysRevA.106.042806}
}

@article{schriber_adaptive_ci_2019,
  author  = {Schriber, Jeffrey B. and Evangelista, Francesco A.},
  title   = {Adaptive Configuration Interaction for Strongly Correlated Electrons with Dynamic Selection of the Active Space},
  journal = {The Journal of Chemical Physics},
  year    = {2019},
  volume  = {151},
  pages   = {174112}
}

@article{herbert_zhu_xray_tddft_2023,
  author  = {Herbert, John M. and Zhu, Ying and Alam, Bushra and Ojha, Avik Kumar},
  title   = {Time-Dependent Density Functional Theory for X-Ray Absorption Spectra: Comparing the Real-Time Approach to Linear Response},
  journal = {Journal of Chemical Theory and Computation},
  year    = {2023},
  volume  = {19},
  number  = {19},
  pages   = {6745-6760},
  doi     = {10.1021/acs.jctc.3c00673},
  URL     = {https://doi.org/10.1021/acs.jctc.3c00673},
  eprint  = {https://doi.org/10.1021/acs.jctc.3c00673}
}

@article{pabst_tdci_2012,
  author  = {Pabst, Stefan and Santra, Robin},
  title   = {Ultrafast and Intense Light-Matter Interactions: Time-Dependent Configuration-Interaction Singles Approach},
  journal = {Physical Review A},
  year    = {2012},
  volume  = {86},
  number  = {6},
  pages   = {063411},
  doi     = {10.1103/PhysRevA.86.063411},
  URL     = {https://doi.org/10.1103/PhysRevA.86.063411},
  eprint  = {https://doi.org/10.1103/PhysRevA.86.063411}
}

@article{goings2018real,
  title={Real-time time-dependent electronic structure theory},
  author={Goings, Joshua J and Lestrange, Patrick J and Li, Xiaosong},
  journal={Wiley Interdisciplinary Reviews: Computational Molecular Science},
  volume={8},
  number={1},
  pages={e1341},
  year={2018},
  publisher={Wiley Online Library}
}

@article{li2020real,
  title={Real-time time-dependent electronic structure theory},
  author={Li, Xiaosong and Govind, Niranjan and Isborn, Christine and DePrince III, A Eugene and Lopata, Kenneth},
  journal={Chemical Reviews},
  volume={120},
  number={18},
  pages={9951--9993},
  year={2020},
  publisher={ACS Publications}
}

@article{terachem21,
author = {Seritan, Stefan and Bannwarth, Christoph and Fales, Bryan S. and Hohenstein, Edward G. and Isborn, Christine M. and Kokkila-Schumacher, Sara I. L. and Li, Xin and Liu, Fang and Luehr, Nathan and Snyder Jr., James W. and Song, Chenchen and Titov, Alexey V. and Ufimtsev, Ivan S. and Wang, Lee-Ping and Martínez, Todd J.},
title = {TeraChem: A graphical processing unit-accelerated electronic structure package for large-scale ab initio molecular dynamics},
journal = {WIREs Computational Molecular Science},
volume = {11},
number = {2},
pages = {e1494},
doi = {https://doi.org/10.1002/wcms.1494},
url = {https://wires.onlinelibrary.wiley.com/doi/abs/10.1002/wcms.1494},
eprint = {https://wires.onlinelibrary.wiley.com/doi/pdf/10.1002/wcms.1494},
year = {2021}
}

@article{Ufimtsev2009b,
Author = {Ufimtsev, Ivan S. and Martinez, Todd J.},
Title = {Quantum Chemistry on Graphical Processing Units. 3. Analytical Energy
   Gradients, Geometry Optimization, and First Principles Molecular
   Dynamics},
Journal = {Journal of Chemical Theory and Computation},
Year = {2009},
Volume = {5},
Number = {10},
Pages = {2619-2628},
Month = {OCT},
DOI = {10.1021/ct9003004},
ISSN = {1549-9618},
EISSN = {1549-9626},
ResearcherID-Numbers = {Martinez, Todd J/F-4306-2010
   Martinez, Todd/HDN-6985-2022},
ORCID-Numbers = {Martinez, Todd J/0000-0002-4798-8947
   },
Unique-ID = {WOS:000270595800005},
}

@article{ Ufimtsev2008,
Author = {Ufimtsev, Ivan S. and Martinez, Todd J.},
Title = {Quantum chemistry on graphical processing units.: 1.: Strategies for
   two-electron integral evaluation},
Journal = {Journal of Chemical Theory and Computation},
Year = {2008},
Volume = {4},
Number = {2},
Pages = {222-231},
Month = {FEB},
DOI = {10.1021/ct700268q},
ISSN = {1549-9618},
EISSN = {1549-9626},
ResearcherID-Numbers = {Martinez, Todd J/F-4306-2010
   Martinez, Todd/HDN-6985-2022},
ORCID-Numbers = {Martinez, Todd J/0000-0002-4798-8947
   },
Unique-ID = {WOS:000253166000002},
}

@article{ Ufimtsev2009a,
Author = {Ufimtsev, Ivan S. and Martinez, Todd J.},
Title = {Quantum Chemistry on Graphical Processing Units. 2. Direct
   Self-Consistent-Field Implementation},
Journal = {Journal of Chemical Theory and Computation},
Year = {2009},
Volume = {5},
Number = {4},
Pages = {1004-1015},
Month = {APR},
DOI = {10.1021/ct800526s},
ISSN = {1549-9618},
EISSN = {1549-9626},
ResearcherID-Numbers = {Martinez, Todd J/F-4306-2010
   Martinez, Todd/HDN-6985-2022},
ORCID-Numbers = {Martinez, Todd J/0000-0002-4798-8947
   },
Unique-ID = {WOS:000265268800039},
}

@article{ durden2022,
Author = {Durden, Andrew S. and Levine, Benjamin G.},
Title = {Floquet Time-Dependent Configuration Interaction for Modeling Ultrafast
   Electron Dynamics},
Journal = {Journal of Chemical Theory and Computation},
Year = {2022},
Volume = {18},
Number = {2},
Pages = {795-806},
Month = {FEB 8},
DOI = {10.1021/acs.jctc.1c01009},
ISSN = {1549-9618},
EISSN = {1549-9626},
ORCID-Numbers = {Levine, Benjamin/0000-0002-0356-0738},
Unique-ID = {WOS:000763089700017},
}

@article{Fales2015,
author = {Fales, B. Scott and Levine, Benjamin G.},
title = {Nanoscale Multireference Quantum Chemistry: Full Configuration Interaction on Graphical Processing Units},
journal = {Journal of Chemical Theory and Computation},
volume = {11},
number = {10},
pages = {4708-4716},
year = {2015},
doi = {10.1021/acs.jctc.5b00634},
URL = {https://doi.org/10.1021/acs.jctc.5b00634}
}

@article{manni2023openmolcas,
  title={The OpenMolcas Web: A community-driven approach to advancing computational chemistry},
  author={Manni, Giovanni Li and Galv{\'a}n, Ignacio Fdez and Alavi, Ali and Aleotti, Flavia and Aquilante, Francesco and Autschbach, Jochen and Avagliano, Davide and Baiardi, Alberto and Bao, Jie J and Battaglia, Stefano and others},
  journal={Journal of Chemical Theory and Computation},
  volume={19},
  number={20},
  pages={6933},
  year={2023}
}

@article{srsen2020limits,
  title={Limits of the nuclear ensemble method for electronic spectra simulations: Temperature dependence of the (E)-azobenzene spectrum},
  author={Srsen, Stepan and Sita, Jaroslav and Slavicek, Petr and Lad{\'a}nyi, V{\'\i}t and Heger, Dominik},
  journal={Journal of Chemical Theory and Computation},
  volume={16},
  number={10},
  pages={6428--6438},
  year={2020},
  publisher={ACS Publications}
}

@article{besley2009time,
  title={Time-dependent density functional theory calculations of near-edge X-ray absorption fine structure with short-range corrected functionals},
  author={Besley, Nicholas A and Peach, Michael JG and Tozer, David J},
  journal={Physical Chemistry Chemical Physics},
  volume={11},
  number={44},
  pages={10350--10358},
  year={2009},
  publisher={Royal Society of Chemistry}
}

@article{calegari2014ultrafast,
  title={Ultrafast electron dynamics in phenylalanine initiated by attosecond pulses},
  author={Calegari, Francesca and Ayuso, D and Trabattoni, Andrea and Belshaw, Louise and De Camillis, Simone and Anumula, S and Frassetto, F and Poletto, Luca and Palacios, Alicia and Decleva, Piero and others},
  journal={Science},
  volume={346},
  number={6207},
  pages={336--339},
  year={2014},
  publisher={American Association for the Advancement of Science}
}

@book{scudder2023electron,
  title={Electron Flow in Organic Chemistry: A Decision-Based Guide to Organic Mechanisms},
  author={Scudder, Paul H},
  year={2023},
  publisher={John Wiley \& Sons}
}

@article{fohlisch2005direct,
  title={Direct observation of electron dynamics in the attosecond domain},
  author={F{\"o}hlisch, Alexander and Feulner, Peter and Hennies, F and Fink, A and Menzel, Dietrich and S{\'a}nchez-Portal, Daniel and Echenique, Pedro M and Wurth, Wilfried},
  journal={Nature},
  volume={436},
  number={7049},
  pages={373--376},
  year={2005},
  publisher={Nature Publishing Group UK London}
}

@article{nisoli2017attosecond,
  title={Attosecond electron dynamics in molecules},
  author={Nisoli, Mauro and Decleva, Piero and Calegari, Francesca and Palacios, Alicia and Mart{\'\i}n, Fernando},
  journal={Chemical Reviews},
  volume={117},
  number={16},
  pages={10760--10825},
  year={2017},
  publisher={ACS Publications}
}

@article{marante2014hybrid,
  title={Hybrid Gaussian--B-spline basis for the electronic continuum: Photoionization of atomic hydrogen},
  author={Marante, Carlos and Argenti, Luca and Mart{\'\i}n, Fernando},
  journal={Physical Review A},
  volume={90},
  number={1},
  pages={012506},
  year={2014},
  publisher={APS}
}

@article{krause2014strong,
  title={Strong field ionization rates simulated with time-dependent configuration interaction and an absorbing potential},
  author={Krause, Pascal and Sonk, Jason A and Schlegel, H Bernhard},
  journal={The Journal of Chemical Physics},
  volume={140},
  number={17},
  year={2014},
  publisher={AIP Publishing}
}

@article{besley2021modeling,
  title={Modeling of the spectroscopy of core electrons with density functional theory},
  author={Besley, Nicholas A},
  journal={Wiley Interdisciplinary Reviews: Computational Molecular Science},
  volume={11},
  number={6},
  pages={e1527},
  year={2021},
  publisher={Wiley Online Library}
}

@article{kochetov2021rhodyn,
  title={RhoDyn: A $\rho$-TD-RASCI framework to study ultrafast electron dynamics in molecules},
  author={Kochetov, Vladislav and Bokarev, Sergey I},
  journal={Journal of Chemical Theory and Computation},
  volume={18},
  number={1},
  pages={46--58},
  year={2021},
  publisher={ACS Publications}
}

@article{lunnemann2009ultrafast,
  title={Ultrafast electron dynamics following outer-valence ionization: The impact of low-lying relaxation satellite states},
  author={L{\"u}nnemann, Siegfried and Kuleff, Alexander I and Cederbaum, Lorenz S},
  journal={The Journal of Chemical Physics},
  volume={130},
  number={15},
  year={2009},
  publisher={AIP Publishing}
}

@article{ruberti2023advances,
  title={Advances in modeling attosecond electron dynamics in molecular photoionization},
  author={Ruberti, Marco and Averbukh, Vitali},
  journal={Wiley Interdisciplinary Reviews: Computational Molecular Science},
  volume={13},
  number={5},
  pages={e1673},
  year={2023},
  publisher={Wiley Online Library}
}

@article{guo2024experimental,
  title={Experimental demonstration of attosecond pump--probe spectroscopy with an X-ray free-electron laser},
  author={Guo, Zhaoheng and Driver, Taran and Beauvarlet, Sandra and Cesar, David and Duris, Joseph and Franz, Paris L and Alexander, Oliver and Bohler, Dorian and Bostedt, Christoph and Averbukh, Vitali and others},
  journal={Nature Photonics},
  volume={18},
  number={7},
  pages={691--697},
  year={2024},
  publisher={Nature Publishing Group UK London}
}

@article{carter2022choice,
  title={On the choice of reference orbitals for linear-response calculations of solution-phase K-edge X-ray absorption spectra},
  author={Carter-Fenk, Kevin and Head-Gordon, Martin},
  journal={Physical Chemistry Chemical Physics},
  volume={24},
  number={42},
  pages={26170--26179},
  year={2022},
  publisher={Royal Society of Chemistry}
}

\end{document}